\documentstyle[12pt]{article}
\def\R{{\rm I\hspace{-.15em}R}}

\def\1{\mbox{I\hspace{-.15em}1}}

\def\b{\begin{equation}}
\def\e{\end{equation}}

\begin{document}

\title{Abelian Gauge Theory in de Sitter Space}
\author{
S. Rouhani $^{1}$ and M.V. Takook$^{2}$
\thanks{e-mail: takook@razi.ac.ir}}

\maketitle \centerline{\it $^{1}$Plasma Physics Research Center,
Islamic Azad University,}
 \centerline{\it P.O.BOX 14835-157, Teheran,
IRAN} \centerline{\it  $^{2}$Department of Physics, Razi
University, Kermanshah, IRAN} \vspace{15pt}

\begin{abstract}

Quantization of spinor and vector free fields in $4$-dimensional
de Sitter space-time, in the ambient space notation, has been
studied in the previous works. Various two-points functions for
the above fields are presented in this paper. The interaction
between the spinor field and the vector field is then studied by
the abelian gauge theory. The $U(1)$ gauge invariant spinor field
equation is obtained in a coordinate independent way notation and
their corresponding conserved currents are computed. The solution
of the field equation is obtained by use of the perturbation
method in terms of the Green's function. The null curvature limit
is discussed in the final stage.

\end{abstract}

\vspace{0.5cm} {\it Proposed PACS numbers}: 04.62.+v, 98.80.Cq,
12.10.Dm \vspace{0.5cm}

\newpage

\section{Introduction}

Quantum field theory in de Sitter space-time has evolved as an
exceedingly important subject, studied by many authors in the
course of the past decade. Historically the de Sitter space-time,
with maximum symmetry in the curved space-time manifold, was
introduced as a solution of the positive cosmological Einstein's
equations. It has the same degree of symmetry as the flat
Minkowski space solution \cite{bida}. The interest in the de
Sitter space increased tremendously when it turned out that it
could play a central role in the inflationary cosmological
paradigm \cite{li}. Very recently, a non-zero cosmological
constant has been proposed to explain the luminosity observations
of the farthest supernovae \cite{pe}. If this hypothesis is
validated in the future, our ideas on the large-scale universe
needs to be changed and the de Sitter metric will play a further
important role.

All these developments make it more compelling than ever to find a
formulation of de Sitter quantum field theory with the same level
of completeness and rigor as for its Minkowskian counterpart. In
Minkowski space, a unique Poincar\`e invariant vacuum can be fixed
by imposing the positive energy condition. In curved space-time,
however, a global time-like Killing vector field does not exist
and therefore the positive energy condition cannot be imposed.
Thus symmetry alone is not sufficient for determination of a
suitable vacuum state. In de Sitter space, however, symmetries
identify the vacuum only in relation to a two parameter ambiguity
$|\alpha,\beta\rangle$, corresponding to a family of distinct de
Sitter invariant vacuum states (see \cite{eila} and references
there in). Only the one parameter family $|\alpha,0\rangle$, is
invariant under the CPT transformation \cite{al,golo,spvo}. By
imposing the condition that in the null curvature limit, the
Wightman two-point function become exactly the same as Minkowskian
Wightman two-point function, the other parameter $(\alpha)$ can be
fixed as well. This vacuum, $|0,0\rangle$, is called Euclidean
vacuum or Bunch-Davies vacuum. It should be noted that this
condition is different with the Hadamard condition, which requires
that the leading short distance singularity in the Hadamard
function $G^{(1)}$ should take its flat space value. The leading
singularity of the Hadamard function is $\cosh 2\alpha$ times its
flat space value.

Bros et al. \cite{brgamo,brmo} presented a QFT of scalar field in
de Sitter space that closely mimics QFT in Minkowski space. They
have introduced a new version of the Fourier-Bros transformation
on the hyperboloid \cite{brmo2}, which allows us to completely
characterize the Hilbert space of ``one-particle'' states and the
corresponding irreducible unitary representations of the de Sitter
group. In this construction the correlation functions are boundary
values of analytical functions. It should be noted that the
analyticity condition is only preserved in the Euclidean vacuum.
In a series of papers we generalized the Bros construction to the
quantization of the various spin free fields in de Sitter space
\cite{ta,ta1,brgamota,gata,gagata1,gagarota,gagata,ta2,morota,paruta}.

In the case of the interaction fields, the tree-level scattering
amplitudes of the scalar field, with one graviton exchange, has
been calculated in de Sitter space \cite{rota}. Recently, the
electromagnetic classical fields produced by the geodesic and
uniformly accelerated discrete charges in de Sitter space-time
have been constructed by Bicak and Krtous \cite{bikr,bikr2}. In
this paper, we show that the $U(1)$ gauge theory in
$4$-dimensional de Sitter space describe the interaction between
the spinor field (``electron'') and the massless vector fields
(``photon''). The ambient space notation {\it i.e.} a coordinate
independent way, has been used throughout this study.

In section $2$, we briefly recall the ambient space notation and
the analysis of the quantum free spinor field
\cite{ta,ta1,brgamota}, and the quantum free vector field
\cite{ta,gata,gagarota} in de Sitter space. Various two-point
functions for free spinor and vector fields in de Sitter space are
then presented. In section $3$, the $U(1)$ gauge invariant spinor
field equation is obtained in the ambient space notation. Their
corresponding conserved current is calculated in a coordinate
independent way. The solution of the field equation is obtained by
the use of the perturbation method. It has been shown that in the
ambient space notation, the formalism of the quantum field in de
Sitter space is very similar to the quantum field formalism in
Minkowski space. The null curvature limit is discussed in the
final stage. A brief discussion in section $4$ concludes this
paper.

\section{Quantum free fields }

\subsection{Spinor fields }

de Sitter space-time is visualized as the hyperboloid with
equation: $$ X_H=\{x^\alpha \in
\R^5:x.x=\eta_{\alpha\beta}x^\alpha
x^\beta=(x^0)^2-(x^1)^2-(x^2)^2-(x^3)^2-(x^4)^2 =-H^{-2}\}$$ \b
\eta^{\alpha\beta}=\mbox{diag}(1,-1,-1,-1,-1);
\;\;\alpha,\beta=0,1,...,4.\e The kinematical group of the de
Sitter space-time is $G_H=SO_0(1,4)$. In this space, we need five
$\gamma$ matrices instead of the usual four in Minkowski
space-time. They are defined by the Clifford algebra:$$ \{
\gamma^\alpha ,\gamma^\beta \}=\gamma^\alpha
\gamma^\beta+\gamma^\beta \gamma^\alpha =
 2 \eta^{\alpha \beta}\;\;\;,\;\;\;
\gamma^{\alpha\dag}=\gamma^0 \gamma^\alpha \gamma^0.$$ An explicit
quaternion representation, which is suitable for symmetry
consideration, is provided by \cite{ta,ta1}
$$ \gamma^0=\left( \begin{array}{clcr} \1 & \;\;0 \\ 0 &-\1 \\ \end{array} \right)
 ,\gamma^4=\left( \begin{array}{clcr} 0 & \1 \\ -\1 &0 \\ \end{array} \right)  $$
  \b   \gamma^1=\left( \begin{array}{clcr} 0 & i\sigma^1 \\ i\sigma^1 &0 \\    \end{array} \right)
     ,\gamma^2=\left( \begin{array}{clcr} 0 & -i\sigma^2 \\ -i\sigma^2 &0 \\  \end{array} \right)
      , \gamma^3=\left( \begin{array}{clcr} 0 & i\sigma^3 \\ i\sigma^3 &0 \\   \end{array} \right)\e
in terms of the $ 2\times2 $ unit $ \1 $ and Pauli matrices
$\sigma^i $. Starting from the Casimir operator and using the
infinitesimal generators, the Casimir eigenvalue equation and some
algebraic relation  dS-Dirac field equation is obtained
\cite{brgamota,dir}\b (-i\not x\gamma.\bar{\partial}
+2i+\nu)\psi(x)=0,\;\; \nu \in \R,\;\;\not x=x.\gamma, \e where
$\bar
\partial_\alpha=\theta_{\alpha \beta}\partial^\beta=
\partial_\alpha+H^2x_\alpha(x. \partial)$. dS-Dirac plane
waves solutions are \cite{ta,ta1,brgamota}  $$ \psi_1^{\xi,{\cal
V}}(x)=(Hx.\xi )^{-2+ i \nu}{\cal V}(x,\xi),$$ $$
\psi_2^{\xi,{\cal U}}(x)=(Hx.\xi)^{-2- i \nu}{\cal U}(\xi),$$
where ${\cal V}$ and ${\cal U}$ are the polarization spinors and
$$\xi \in {\cal C}^+=\{ \xi \;\;;\eta_{\alpha \beta}\xi^\alpha
\xi^\beta=(\xi^0)^2-\vec \xi.\vec \xi-(\xi^4)^2=0,\; \xi^0>0 \}.
$$ Due to the singularity and sign phase ambiguity, the solution
are  defined in the complex de Sitter space \cite{brgamo}, $$z \in
X_H^{(c)}\equiv \{ z=x+iy\in C^5 ;\;\;\eta_{\alpha \beta}z^\alpha
z^\beta=(z^0)^2-\vec z.\vec z-(z^4)^2=-H^{-2}\}, $$ $$
\psi_1^{\xi,{\cal V}}(z)=(Hz.\xi )^{-2+ i \nu}{\cal V}(z,\xi),$$
$$ \psi_2^{\xi,{\cal U}}(z)=(Hz.\xi)^{-2- i \nu}{\cal U}(\xi).$$
The spinor field operator is defined by the boundary value of
complex solutions $$ \psi(x)=\int_T \sum_{a=1,2} \{\;
a_a(\xi,\nu){\cal
U}^a(\xi)[(x.\xi)_+^{-2-i\nu}+e^{-i\pi(-2-i\nu)}(x.\xi)_-^{-2-i\nu}]
$$ \b +d^{\dag}_a(\xi,\nu)H\not x\not {\xi}{\cal
V}^a(\xi)[(x.\xi)_+^{-2+i\nu}+e^{-i\pi(-2+i\nu)}(x.\xi)_-^{-2+i\nu}]\;
\} d\mu_T(\xi),\e where $T$ denotes an orbital basis of ${\cal
C}^+$ and \cite{geshi} $$  (x\cdot
\xi)_+=\left\{\begin{array}{clcr} 0 & \mbox{for} \; x\cdot \xi\leq
0,\\ (x\cdot \xi) & \mbox{for} \;x\cdot \xi>0. \\ \end{array}
\right. $$  $ d\mu_T(\xi)$ is an invariant measure on ${\cal C}^+$
\cite{brmo}. The vacuum state, which is fixed by imposing the
condition that in the null curvature limit the Wightman two-point
function become exactly the same as Minkowskian Wightman two-point
function, is defined as follows
$$a_a(\xi,\nu)|\Omega>=0=d_a(\xi,\nu)|\Omega>.$$ This vacuum, $|\Omega>$, is
equivalent to the Euclidean  vacuum $|0,0\rangle$. ``One particle
'' and ``one anti-particle'' states are \b
d^{\dag}_a(\xi,\nu)|\Omega>=|\xi,a,\nu>,\;\;\;a^{\dag}_a(\xi,\nu)|\Omega>=\overline{|\xi,a,\nu>}.\e

Various two-point functions for scalar field can be calculated in
terms of the Wightman two point function, ${\cal
W}^{\nu}({x},{x'})$ \cite{brmo},
\begin{itemize}  \item{ commutator function}
  $$  {\cal C}^{\nu}({x},{x'})={\cal W}^{\nu}({x},{x'})-{\cal W}^{\nu}({x'},{x}),$$
\item{ retarded propagator }
    $$ G_r^{\nu}({x},{x'})=i \theta(x^{(0)}-x'^{(0)}){\cal C}^{\nu}({x},{x'}), $$
\item{ advanced propagator }
   $$G_a^{\nu}({x},{x'})=G_r^{\nu}({x},{x'})-i{\cal C}^{\nu}({x},{x'}),$$
\item{ chronological propagator }
   $${\cal T}^{\nu}({x},{x'})=-iG_r^{\nu}({x},{x'})+{\cal W}'^{\nu}({x},{x'})=
   -iG_a^{\nu}({x},{x'})+{\cal W}^{\nu}({x},{x'}),$$
\item{ anti-chronological propagator }
   \b\bar {{\cal T}}^{\nu}({x},{x'})=iG_r^{\nu}({x},{x'})+{\cal W}^{\nu}({x},{x'})=
   iG_a^{\nu}({x},{x'})+{\cal W}'^{\nu}({x},{x'}),\e

\end{itemize}
where all above functions have been presented in the coordinate
independent way notation. For the spinor field, the matrix
Wightman two point function is \cite{ta,ta1,brgamota}
$${\cal W}^{\nu}({x},{x'})=<\Omega,\psi(x)\bar \psi(x') \Omega
>=D(x',\partial_{x'}){\cal
W}_{0}({x},{x'})= \frac{i\nu(1+\nu^2)}{64\pi \sinh(\pi \nu)}
\times $$
$$\left[ (2-i\nu)P^{(7)}_{-2-i\nu}\! ({\cal Z}(x,x')){\not x}-
(2+i\nu)P^{(7)}_{-2+i\nu}\! ({\cal Z}(x,x')){\not x}'\right] \!
\gamma^{4},$$ \vskip 0.5 cm \noindent where $\bar
\psi=\psi^\dag\gamma^0\gamma^4$. ${\cal W}_{0}({x},{x'})$ is the
two-point function of the scalar field and $D(x',\partial_{x'})$
is the matrix differential operator \cite{ta} \b
D(x',\partial_{x'})=\frac{1}{\nu+i}(-i\not
x'\not{\bar{\partial}}_{x'}+i+\nu).\e  $P^{(7)}_{-2-i\nu}\! ({\cal
Z}(x,x')) $ is a generalized Legendre function \cite{brmo} and
$${\cal Z}(x,x')=-H^2 x.x'=1+\frac{H^2}{2} (x-x')^2 \equiv \cosh H
\sigma (x,x'). $$

Similar to the case of Minkowski space \cite{boloto}, various
two-point functions for spinor field are obtained from the
corresponding functions of the scalar field by: \b {\cal
S}_\nu^{Z}(x,x')=D(x',\partial_{x'}){\cal W}_{0}^Z({x},{x'}),\e
where ${\cal W}_{0}^Z$ are the various two-point functions for
scalar field eq. $(6)$. $Z$ stands for retarded, advanced,
chronological and anti-chronological propagators. For example, the
anti-commutator function is given by \cite{brgamota}
$${\cal A}^{\nu}({x},{x'})=<\Omega,\{\psi(x) \bar \psi(x')\}
\Omega >= \frac{i\nu(1+\nu^2)}{32\pi R^2}
\epsilon(x^{0}-x'^{0})\theta({\cal Z}-1) \times $$ $$\left[
(2-i\nu)P^{(7)}_{-2-i\nu}\! ({\cal Z}(x,x')){\not x}+
(2+i\nu)P^{(7)}_{-2+i\nu}\! ({\cal Z}(x,x')){\not x}'\right] \!
\gamma^{4},$$ \vskip 0.5 cm \noindent where $\theta$ is the
Heaviside step function and $$ \epsilon
(x^0,x'^0)=\left\{\begin{array}{clcr} 1&x^0>x'^0
\\ 0&x^0=x'^0\\ -1&x^0<x'^0.\\   \end{array} \right.$$

\subsection{Vector fields }

Starting from the Casimir operator and using the infinitesimal
generators, the Casimir eigenvalue equation and some algebraic
relation the field equation for a massless vector field in the
ambient space notation is obtained \cite{ta,gata,gagarota} \b
(H^{-2}(\bar
\partial)^2+2)K(x)-2x\bar
\partial.K(x)-H^{-2}\bar \partial \partial.K=0.\e
This five-component vector field quantity has to be viewed as a
homogeneous function of the $\R^5$-variables $x^{\alpha}$ with
some arbitrarily chosen degree $\sigma$ \cite{dir}
        \b x^{\alpha}\frac{\partial }{\partial
x^{\alpha}}K_{\beta}(x)=x\cdot \partial K_\beta (x)=\sigma
        K_{\beta}(x). \e
The direction of $K_\alpha(x)$ lies in the de Sitter space if we
require the condition of transversality $ x\cdot K(x)=0 $. The
above field equation is gauge invariant, {\it i.e.}
   \b K\longrightarrow K'=K+H^{-2}\bar \partial \phi_g \e
\b(H^{-2}(\bar
\partial)^2+2)K'(x)-2x\bar
\partial.K'(x)-H^{-2}\bar \partial \partial.K'(x)=0, \e where $\phi_g$ is an
arbitrary scalar field. Similar to  the flat space massless vector
field, the gauge fixing is accomplished by adding to $(9)$ a gauge
fixing term. We obtain \cite{ta,gagarota}\b (H^{-2}(\bar
\partial)^2+2)K(x)-2x\bar
\partial.K(x)-c H^{-2}\bar \partial \partial.K=0,\e
where $c$ is the gauge fixing parameter. The simple choice for $c$
is $\frac{2}{3}$. The plane wave solutions in this gauge, are
\cite{ta,gagarota} $$ K_1^{\xi,\lambda}(z)={\cal
E}_{1}^{(\lambda)}(z,\xi)(Hz.\xi)^{-1},$$
  \b K_2^{\xi,\lambda}(z)={\cal
   E}_{2}^{(\lambda)}(z,\xi)(Hz.\xi)^{-2},\e where ${\cal
E}_{1,2}^{(\lambda)}(z,\xi)$ are the polarization vectors. In this
gauge, we can write the ''massless'' vector field in terms of the
polarization vectors and a ''massless'' conformally coupled scalar
field.

In this gauge, the Wightman vector two-point function is
\cite{ta,gagarota}$$ {\cal W}_{\alpha \alpha'}(x,x')=\langle
\Omega,K_{\alpha}(x)K_{\alpha'}(x')\Omega \rangle= D_{\alpha
\alpha'}(x,x',\partial_{x})W_{0}(x,x'),$$ where $W_{0}(x,x')$ is
the Wightman two-point function of the conformally coupled scalar
field \cite{tag,ta} \b  W_{0}(x,x')=
\frac{-1}{8\pi^2}\left[\frac{H^2}{1-{\cal Z}(x,x')}-i\pi
H^2\epsilon(x^0-x'^0)\delta(1-{\cal Z}(x,x'))\right],\e and
$D_{\alpha \alpha'}(x,x',\partial_{x})$ are the bi-tensor
differential operator
 \b D_{\alpha \alpha'}(x,x',\partial_{x})= \theta_{\alpha }.\theta'_{\alpha' }+H^{-2} \bar
\partial_\alpha \left[\bar
\partial . \theta'_{\alpha' }+H^2x.\theta'_{\alpha'
}\right].\e

Similar to the spinor field, the various two point functions for
massless vector field are obtained from the corresponding
functions of the scalar field according to the formula \b
G_{\alpha \alpha'}^{Z}(x,x')=D_{\alpha
\alpha'}(x,x',\partial_{x})G_{0}^Z(x,x').\e where $G_{0}^Z$ stands
for the two point functions of scalar fields eq. $(6)$. For
example, the commutator function is $$ G_{\alpha \alpha'}(x,x')=
\langle \Omega,[K_{\alpha}(x)K_{\alpha'}(x')]\Omega
\rangle=D_{\alpha \alpha'}(x,x',\partial_{x})G_{0}(x,x')$$ \b =
\left( \theta_{\alpha }.\theta'_{\alpha' }+H^{-2} \bar
\partial_\alpha \left[\bar
\partial . \theta'_{\alpha' }+H^2x.\theta'_{\alpha'
}\right]\right) \frac{-H^2}{8\pi}\epsilon(x^0-x'^0)\delta(1-{\cal
Z}(x,x')),\e Where $G_{0}(x,x')$ is the commutator function of the
conformally coupled scalar field \cite{tag,ta}.

\section{Gauge invariant field equation}

dS-Dirac field equation is invariant under the $U(1)$ global
symmetry, \b \psi(x) \longrightarrow \psi'(x)=e^{-i\Lambda}\psi(x)
\e $$ (-i\not x\gamma.\bar{\partial} +2i+\nu)\psi'(x)=0,$$ where
$\Lambda$ is a constant. This equation is not invariant under the
$U(1)$ local gauge symmetry \b \psi(x) \longrightarrow
\psi'(x)=e^{-i\Lambda(x)}\psi(x), \e \b (-i\not
x\gamma.\bar{\partial} +2i+\nu)\psi'(x)\neq 0.\e The notation of
local gauge symmetry with its space-time-dependent transformation
can be used to generate the gauge interaction. The abelian $U(1)$
local symmetry is defined by the interaction between the electron
field and the electromagnetic field {\it i.e.} ``electron-photon''
interaction.

For obtaining a dS-Dirac local gauge invariant equation, it is
necessary to replace the covariant derivative
$\bar{\partial}_\beta$ with the gauge-covariant derivative
$D_\beta$ which is defined by $$ D_\beta =\bar{\partial}_\beta
+iqA_{\beta}, $$ where $A_{\beta}$ is a new vector field or the
gauge field, and $q$ is a free parameter which can be identifies
with the electric charge in the null curvature limit. If the gauge
field $A_{\beta}(x)$ has the transformation property \b
A_\beta\longrightarrow A'_\beta=A_\beta+\frac{1}{q}\bar
\partial_\beta \Lambda (x) ,\e the gauge-covariant derivative of the spinor
field has the following simple transformation \b D_\beta\psi(x)
\longrightarrow [D_\beta
\psi(x)]'=e^{-i\Lambda(x)}D_\beta\psi(x),\e  Comparting equation
$(22)$ with the equation $(11)$, the simplest choice of $A_\beta$
(gauge-invariant vector field) is the massless vector field
$K(x)$, which was presented in the previous section. Thus the
dS-Dirac local gauge invariant equation is $$ (-i\not x\gamma.D
+2i+\nu)\psi(x)=0, $$ \b (-i\not x\gamma.\bar{\partial}+q\not
x\gamma.K +2i+\nu)\psi(x)=0.\e This equation is simultaneously
invariant under the two transformations $(20)$ and $(22)$
$$(-i\not x\gamma.D' +2i+\nu)\psi'(x)=0,$$ where $$ D_\beta'
=\bar{\partial}_\beta +iqK'_{\beta}.$$

The Noether's theorem states that any lagrangian invariant under a
continuous one-parameter transformation, is associated with a
local conserved current. By using the dS-Dirac equation for $
\psi$ and $ \bar\psi$ \cite{brgamota}, $$  \bar \psi\gamma^4(i
\overleftarrow{\not\bar\partial} \not x-2i+\nu)=0, $$ the free
field lagrangian is defined by $$ {\cal L}_0=H\bar
\psi\gamma^4(-i\not x\gamma.\bar{\partial}+2i+\nu)\psi(x).$$ Thus
the invariance of the free field lagrangian under the
transformation $(19)$ leads to the following conserved current \b
J_\alpha(x )= \frac{\delta {\cal L}_0}{\delta
\bar\partial^{\alpha} \psi} \frac{\delta \psi}{\delta \Lambda
}=-H\bar \psi(x)\gamma^4\not x \gamma_\alpha \psi(x).\e It is
easily verified that this current satisfies the following
conditions
$$\bar\partial.J(x)=0,\;\;\;\; x.J(x)=0.$$ The field equation $(24)$
gives the interaction between the spinor field (electron) and the
massless vector field (photon). We can write the interaction
Lagrangian between electron and photon in the following way $$
{\cal L}_{int}=qH\bar \psi(x)\gamma^4 \not x\gamma.K(x) \psi(x).$$

By choosing  electromagnetic field $K(x)$, we can obtain the
solution of the field equation $(24)$ by perturbation method. It
can be written in the following form \b (-iH\not
x\gamma.\bar{\partial} +2iH+\nu H)\psi(x)=-qH\not x\gamma.K(x)
\psi(x).\e In practice, this equation explains the interaction of
an electron field $\psi(x)$ and an electromagnetic field $K(x)$ in
de Sitter space. Its solution is discussed by the use of the Green
function method. The matrix Green function, is solution of the
following equation $$ (-iH\not x\gamma.\bar{\partial} +2iH+\nu
H){\cal S}^Z_\nu(x,x')=\delta_H(x-x'),$$ where ${\cal
S}^Z_\nu(x,x')$ is retarded or advance Green function, obtained in
the previous section (eq. $(8)$). $\delta_H(x-x')$ is the Dirac
delta function in the ambient space notation on the de Sitter
hyperboloid \b \int d\sigma(x')\delta_H(x-x')f(x')=f(x),\e where
$d\sigma(x')$ is the dS-invariant volume \cite{brmo} and $f(x)$ is
a homogeneous function (with some arbitrarily chosen degree) of
the $\R^5$-variables $x^{\alpha}$. Therefore the solution of the
field equation can be written in the following form $$
\psi(x)=\psi_0(x)+ (-qH)\int d\sigma(x'){\cal S}^\nu(x,x') \not
x'\gamma.K(x') \psi(x'),$$ where $\psi_0(x)$ is a free field
solution $(3)$. In the perturbation theory, solution is $$
\psi(x)=\psi_0(x)+ (-qH)\int d\sigma(x'){\cal S}^\nu(x,x') \not
x'\gamma.K(x') \psi_0(x')$$ \b +(-qH)^2\int
d\sigma(x')d\sigma(x''){\cal S}^\nu(x,x') \not x'\gamma.K(x')
{\cal S}^\nu(x',x'') \not x''\gamma.K(x'') \psi_0(x'')+O(q^3).\e
In this notation, the de Sitter QFT formalism is very similar to
the Minkowskain counterpart.

Finally, we consider the Minkowskian limit $(H=0$). The dS point
$x=x_H(X)$ has been expressed in terms of the Minkowskian variable
$X=(X_0=ct, \vec X)$ measured in units of the dS radius $H^{-1}$:
 $$ x_H(X)=(x^0=H^{-1}\sinh HX^0, \vec x=H^{-1} \frac{\vec X}{\parallel
\vec X\parallel} \cosh HX^0 \sin H\parallel \vec X\parallel,$$ \b
x^4= H^{-1}  \cosh HX^0 \cos H\parallel \vec X\parallel).\e Note
that $(X^0, \vec X)$ are global coordinates. In the null curvature
limit, we have $x^\alpha=(X^0, \vec X,H^{-1} )=(X^\mu, x^4=H^{-1}
).$ In this limit the field equation $(3)$ reads
\cite{ta,brgamota}\b (i \gamma_\mu\gamma^4{\partial}^\mu -q
\gamma_\mu\gamma^4 K^\mu(x)-\nu H)\psi(x)= 0,\e where in the null
curvature limit $\nu H\longrightarrow m$ and $ q\longrightarrow
e$. They are mass and electric charge of the electron,
respectively, in the Minkowski space. $ \gamma^\mu
\gamma^4=\gamma'^\mu$ are the usual Dirac matrices in the
Minkowski space. This is exactly the Dirac equation in the
presence of an electromagnetic field in Minkowski space \b (i
\gamma'_\mu{\partial}^\mu -e \gamma'_\mu A^\mu(X)-m)\psi(X)= 0.\e

\section{ Conclusions}

The formalism of the quantum field in de Sitter universe, in
ambient space notation, is very similar to the quantum field
formalism in Minkowski space. The Fourier-Bros transformation on
the hyperboloid allows us to write the fields in terms of two
separate parts, a polarized and ``plane wave'' part. In this
notation the concept of the ``particle states'', contrary to the
concept of energy, is defined globally. The Fock space is
constructed in terms of the five-vector $\xi$, which in the
Minkowskian limit is the four energy-momentum vector ($p^\mu$).
The interaction between the quantum charge particle and a
classical electromagnetic field can be described by the
perturbation method in terms of Green function, very similar to
the Minkowski space.

Similarly nonabelian gauge fields can be constructed as well. The
importance of this formalism may be shown further by the
consideration of the linear quantum gravity and supergravity in de
Sitter space, which lays a firm ground for further study of
universe.

\vskip 0.5 cm \noindent {\bf{Acknowledgements}}: The authors would
like to thank Professor J. Bros for providing us with his valuable
paper \cite{brmo2} before submission on WEB. We also extend our
gratitude to S. Moradi for constructive discussions on this
subject.


\begin{thebibliography}{a}
\addcontentsline{toc}{chapter}{Bibliographie}

\bibitem{bida} Birrel N.D. and Davies P.C.W., Cambridge Univ. Press, (1982)
{\it Quantum Fields in Curved Space}
\bibitem{li} Linde A., Harwood Academic Publishers, Chur (1990),
{\it Particle Physics and Inflationary Cosmology}
\bibitem{pe} Perlmutter S. et al, Astrophys. J. 517(1999)565, astro-ph/9812133
\bibitem{eila} Einhorn M.B., Larsen F., Phys. Rev. D
65(2002)104037, hep-th/0112223
\bibitem{al} Allen B., Phys. Rev. D 32(1985)3136
\bibitem{golo} Goldstein K., Lowe D. A., Nucl. Phys. B
669(2003)325, hep-th/0302050
\bibitem{spvo} Spradlin M., Volovich A., Phys. Rev. D, 32(1985)3136
\bibitem{brgamo} Bros J., Gazeau J.P., Moschella U., Phys. Rev. Lett. 73(1994)1746
\bibitem{brmo} Bros B. and Moschella U., Rev. Math. Phys.
8(1996)327, gr-qc/9511019
\bibitem{brmo2} Bros B. and Moschella U., {\it Fourier analysis
and holomorphic decomposition on the one-sheeted hyperboloid},
math-ph/0311052
\bibitem{ta} Takook M.V., Th\`ese de l'universit\'e Paris VI,
(1997)
\bibitem{ta1} Takook M.V., Proceeding of the Group 21, 15-20 july (1996),
gr-qc/0005077
\bibitem{brgamota} Bartesaghi P., Gazeau J-P., Moschella  U. and Takook M.V., Class.
Quantum Grav. 18(2001)4373
\bibitem{gata} Gazeau J. P., Takook M.V., J. Math. Phys. 41(2000)5920, gr-qc/9912080
\bibitem{gagata1} Garidi T., Gazeau J-P. and
Takook M.V., J. Math. Phys. 43(2002)6379
\bibitem{gagarota} Garidi T., Gazeau J-P., Rouhani S. and Takook M.V., {\it ``Massless'' vector field in de Sitter
space}, in preparation
\bibitem{gagata} Garidi T., Gazeau J-P. and Takook M.V., J. Math. Phys. 44(2003)3838,
hep-th/0302022
\bibitem{ta2} Takook M.V., Proceeding of the Wigsym6, July 16-22 August, (1999), Istanbul, Turkey, gr-qc/0001052
\bibitem{morota} Moradi S. et al., Phys.
Lett. B 613(2005)74, gr-qc/0502022
\bibitem{paruta} Pahlavan A. et al., {\it $N=1$ de Sitter Supersymmetry Algebra},
appear in Phys. Lett. B (2005), gr-qc/0506099
\bibitem{rota} Rouhani S. and Takook M.V., EuroPhys. lett. 68(2004)15, gr-qc/0409120
\bibitem{bikr}  Bicak J., Krtous P., Phys. Rev. D 64(2001)124020,
gr-qc/0107078
\bibitem{bikr2}  Bicak J., Krtous P., Phys. Rev. Lett.
88(2002)211101, gr-qc/0207010
\bibitem{dir}  Dirac P. A. M., Annals of Math. 36(1935)657
\bibitem{geshi} Gel'fand I.M. and Shilov E.G., Academic Press (1964), {\it Generalized Functions}
\bibitem{boloto} Bogolubov N. N., Longunov A. A., Todorov I. T.,
B. W. Benjamin Inc. (1975) {\it Intriduction to Axiomatic Quantum
Field Theory}
\bibitem{tag} Tagirov E. A., Annals of Phys. 76(1973)561

\end{thebibliography}
\end{document}